\documentclass[aps,pra,twocolumn,showpacs]{revtex4-1}
\usepackage{graphicx,epstopdf}
\usepackage{color}
\usepackage{float}
\usepackage{epsfig}
\usepackage{epstopdf}
\usepackage{times}
\usepackage{amsmath, amssymb}
\newcommand{\mathsym}[1]{{}}
\newcommand{\unicode}[1]{{}}

\usepackage[colorlinks=true, citecolor=blue, urlcolor=blue ]{hyperref}

\begin{document}

\title{Effects of cavity-cavity interaction on the entanglement dynamics\\ of  
a generalized double Jaynes-Cummings model}

\author{Mahasweta Pandit,\(^{1,2}\) Sreetama Das,\(^{1}\) Sudipto Singha Roy,\(^{1}\) Himadri Shekhar Dhar,\(^{1,3}\)  and Ujjwal Sen\(^{1}\)}

\affiliation{\(^1\)Harish-Chandra Research Institute, HBNI, Chhatnag Road, Jhunsi, Allahabad 211 019, India}
\affiliation{\(^2\)National Institute of Technology, Rourkela, Odisha 769 008, India}
\affiliation{\(^3\)Institute for Theoretical Physics, Vienna University of Technology, Wiedner Hauptstra{\ss}e 8-10/136, A-1040 Vienna, Austria}

\date{\today}

\begin{abstract}
We consider a generalized double Jaynes-Cummings model consisting of two isolated two-level atoms, each contained in a lossless cavity that interacts with each other through a controlled photon-hopping mechanism. We analytically show that at low values of such a mediated cavity-cavity interaction, the temporal evolution of entanglement between the atoms, under the effects of cavity perturbation, exhibits the well-known phenomenon of entanglement sudden death. Interestingly, for moderately large interaction values, a complete preclusion of entanglement sudden death is achieved, irrespective of its value in the initial atomic state. Our results provide a model to sustain entanglement between two atomic qubits, under the adverse effect of cavity induced perturbation, by introducing a non-intrusive inter-cavity photon exchange that can be physically realized through cavity-QED setups in contemporary experiments.
\end{abstract}

\maketitle

\section{Introduction}
Understanding the effects of decoherence on evolution of {a} quantum system is an important aspect  of various investigations in quantum information theory~\cite{op_quant_sys_rev1,op_quant_sys_rev2,op_quant_sys_rev3,op_quant_sys_rev4,op_quant_sys_rev5,
op_quant_sys_rev6,op_quant_sys_rev7,op_quant_sys_rev8,op_quant_sys_rev9,op_quant_sys_rev10,
op_quant_sys_rev11,op_quant_sys_rev12,op_quant_sys_rev13}. Quantum correlation, in particular quantum entanglement~\cite{entanglement}, is extremely fragile to environmental perturbations and exhibits rapid decay against surrounding noises,  often leading to the phenomenon of  entanglement sudden death (ESD)~\cite{SD_main}. Therefore, for practical implementation of different quantum information theoretic protocols such as superdense coding~\cite{DC}, quantum teleportation~\cite{tele}, and quantum key distribution~\cite{QKD}, using composite quantum systems~\cite{exp_many_body1,exp_many_body2,exp_many_body3}, one of  the primary motivations is to examine and counteract the adverse effects of environmental noise in quantum system.

In this work, we consider a generalized double Jaynes-Cummings (JC) model~\cite{JC1,JC2,scully}, consisting of two cavities, each of which contains a single two-level atom inside. Our main aim is to study the dynamics of entanglement between the isolated atoms, under the constraint that energy transfer between the cavities is permissible through a controlled photon-hopping mechanism \cite{qptlight,oztop}. The model  is significant from a  quantum information theoretic perspective as the nonclassical correlations generated due to the interaction between the atoms and the modes of the cavities can be harnessed for performing different quantum information processing (QIP) tasks~\cite{DC,tele,QKD}. The JC model, consisting of a single atom-photon interaction, was first introduced to study spontaneous emission and examine the semiclassical behavior associated with quantum radiation~\cite{scully}.
In the following years, in order to gain a deeper insight of the  underlying phenomenon associated with atom-photon interactions, various generalizations of  the JC model have  been proposed, even  in the multiparty domain~\cite{review_JC,SD1,SD2,SD3,SD4,SD5,SD6,SD7,SD8,new1}. Of particular interest is the case of the double JC model, consisting of two  atoms, each inside a pair of lossless cavities~\cite{SD1}. It was reported that the time-evolving atom-cavity systems, which do not mutually interact, follow an unusual dynamics of entanglement between the atoms, leading to  ESD. Moreover, since the pair of atom-cavity systems form a closed system, the entanglement revives after sudden death in a periodic manner. Interestingly, the time period over which the bipartite entanglement remains zero, after ESD and before revival, turns out to be  an explicit function of the entanglement of the initial two-atom state.
However, this correspondence between the initial entanglement and the duration of death is not valid if only a single cavity is present~\cite{SD2}, and though the system still undergoes ESD, the hierarchy of initial states, based on their entanglement, is lost.
Moreover, it has also been reported that  instead of considering the cavities in their respective ground states, if one starts with two different initial light fields, say the squeezed vacuum state and the coherent state, then the double intensity-dependent JC model~\cite{SD3},  would generate a periodic entanglement pulse which may be useful for QIP protocols.
Further generalization in this direction  incorporates the influences of atom-field coupling and dipole-dipole coupling,  on the entanglement between the isolated atoms, kept in their respective cavities~\cite{SD8}.

Therefore, it is evident from the above examples that in the double JC model, the effects of cavity-induced  perturbations on the dynamics of entanglement leads to an unavoidable sudden death, thus making the bipartite atomic state inefficient for performing various QIP tasks. Hence, there is a major motivation  to design atom-cavity models that can potentially avoid entanglement loss in an efficient manner.
%
In Ref.~\cite{SD_prev2}, it was reported that by choosing different decay coefficients of the atoms from the upper to any other low energy level, the phenomenon of ESD  can be mitigated.
Further proposals in the direction involve sophisticated schemes based on Zeno-like measurements during the dynamical process~\cite{SD6}, resulting in not only the prevention of ESD but also the enhancement of entanglement.

In our work, we attempt to overcome the sudden death of entanglement in a double JC model,
by modelling a photon exchange mechanism between the two cavities, thus introducing a cavity-cavity interaction in the Hamiltonian of the system. This is motivated by the fact that such photon-hopping between cavities is routinely used in the simulation of strongly correlated dynamics on a mesoscopic scale, using arrays of atom-cavity systems \cite{qptlight,oztop}. Moreover, these models can potentially be realized in experiments using, among others, photonic band gap structure of nitrogen-vacancy (NV) centers in diamonds, placed in a cavity \cite{nv} or microwave strip line resonators \cite{strip}.
Based on analytical results, we report that for low values of the cavity-cavity interaction, entanglement between the isolated atoms, as quantified by concurrence~\cite{Concurrence}, evolves in almost a similar fashion as compared to the case when no interaction is present~\cite{SD1}. However, a significantly different behavior can be observed when the cavity-cavity interaction term is moderately increased.  We observe that for these values, a complete prevention of  entanglement loss through ESD can be achieved, irrespective of the entanglement content of the initial state. Moreover, the entanglement between the two cavities remains negligibly small, which is contrary to the  behavior obtained when photon exchange between the cavities is forbidden.
To highlight the importance of our results, we compare the effect of cavity-cavity interaction with the more intrusive case where the atoms directly interact through Ising or $XY$ spin-exchange interactions~\cite{JC_XY}. We observe that for the Ising interaction, a strong spin-spin coupling is required to overcome ESD, whereas,  for the $XY$ interaction, circumventing sudden death is not possible for certain parameter ranges even for high coupling.
%
\section{Methodology} 
\label{model}
The dynamics of the entanglement between the atoms, governed by the double JC  model with lossless non-interacting cavities, is sensitive to cavity induced perturbations and for certain initial states may face sudden death~\cite{SD1}.
This indeed makes  the quantum channel, shared by the atoms, inefficient for performing several QIP tasks.
Therefore, at this stage, one may ask the following question: ``Is there a non-intrusive way to minimize the cavity effects such that ESD can be avoided?''
\begin{figure}[h]
\begin{center}
\includegraphics[scale=.25]{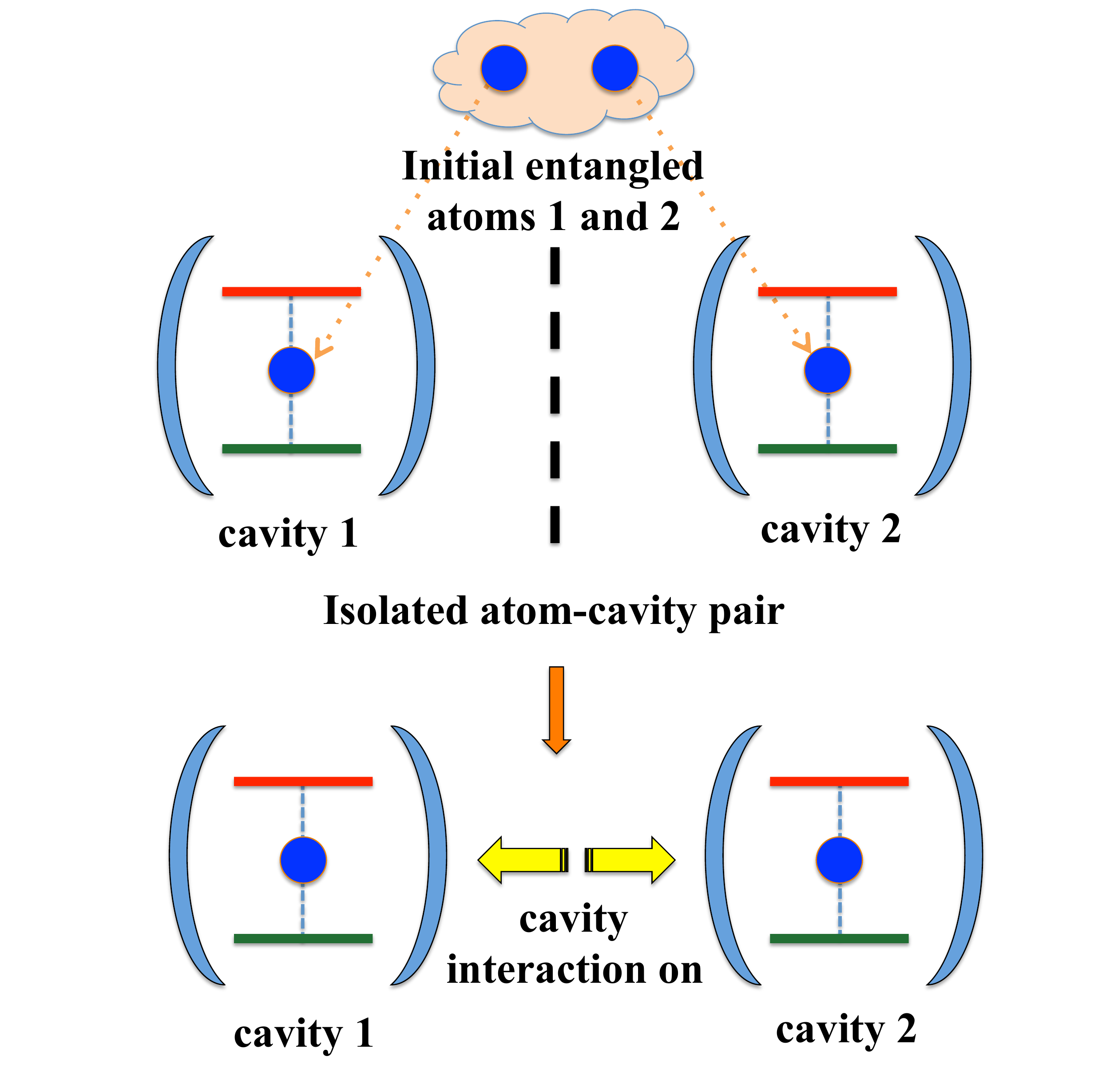}
\caption{(Color online.) A schematic diagram of a double Jaynes-Cummings model comprised of two spatially separated atoms, with some initial entanglement, inside a pair of lossless cavities. The atom is  taken to be a two-level system, with the cavities represented by the Fock basis. The cavity-cavity interaction discussed in this work allows the cavities to exchange photons during the evolution of the system.}
\label{schematic}
\end{center}
\end{figure}
A plausible strategy is to consider a photon-exchange mechanism between the cavities, without directly affecting the atoms in the system. The energy transfer between the cavities may allow sufficient re-coherence between the isolated atoms to bypass sudden death and make the two-atom system more robust against  perturbations induced by the cavities.
From our context,  we first modify  the Hamiltonian, for the two isolated atom-cavity pairs by including a cavity-cavity interaction term, in the following way:
\begin{eqnarray}
\mathcal{H}'&=& \omega \sum_{i=1}^2 \sigma^z_i+g\sum_{i=1}^2 (a^+_i\sigma^-_i+a_i\sigma^+_i) +\nu\sum_{i=1}^2 a_i^+ a_i \nonumber\\
&+& \kappa \sum_{\substack{i,j=1\\ i\neq j}}^2a^+_i a_j,
\label{JC_modified}
\end{eqnarray}
where $\omega_i$ is the transition frequency between the atomic levels $|e_i\rangle$ and $|g_i\rangle$, $\nu$ is the frequency of the harmonic oscillator mode of the cavity, and $g_i$ is the atom-cavity coupling constant, also known as the single-photon Rabi frequency.  $\kappa $ is the cavity-cavity  coupling term.
Such interactions between cavities have been envisaged in the study of strongly correlated arrays of optical cavities \cite{qptlight, oztop}, by introducing photon-hopping between cavities, to simulate Bose-Hubbard models. Moreover, $\sigma^+_i(\sigma^-_i)$ are the atomic raising (lowering) operator for the $i^{th}$ atom, and $a^+_j~ (a_j)$ are the bosonic creation (annihilation) operators for the $j^{th}$ cavity. $\sigma_z$ is the atomic inversion operator given by the Pauli spin operator along the $z$ direction. Here, for all computation purpose we kept $\omega_1=\omega_2=\omega$ and $g_1=g_2=g$.


\section{Effect of  cavity-cavity interactions}
\label{Effect of cavity-cavity interactions}
An important simplification in investigating the dynamics of the double JC model, is to make use of the symmetry. Since, the pair of atom-cavity systems form a closed systems, the total energy in the system is a conserved quantity. Hence, for a maximum of double excitation in the atomic state, the Hamiltonian of the system can be expressed using the following basis states  ($|v_i\rangle$) for the total atom-cavity system:
$\{|\hspace{-0.1cm}\uparrow\uparrow\hspace{-0.1cm}00\rangle,  |
\hspace{-0.1cm}\uparrow\downarrow\hspace{-0.1cm}10\rangle,|
\hspace{-0.1cm}\uparrow\downarrow\hspace{-0.1cm}01\rangle,|
\hspace{-0.1cm}\downarrow\uparrow\hspace{-0.1cm}10\rangle,|
\hspace{-0.1cm}\downarrow\uparrow\hspace{-0.1cm}01\rangle,|
\hspace{-0.1cm}\downarrow\downarrow\hspace{-0.1cm}20\rangle,|
\hspace{-0.1cm}\downarrow\downarrow\hspace{-0.1cm}11\rangle,|
\hspace{-0.1cm}\downarrow\downarrow\hspace{-0.1cm}02\rangle,|
\hspace{-0.1cm}\downarrow\downarrow\hspace{-0.1cm}00\rangle$\}, where $\{|0\rangle, |1\rangle,|2\rangle,..,|n\rangle\}$ are the basis states in the Fock space and $|\uparrow\rangle(|\downarrow\rangle)$ represents atomic excited (ground) state.  The matrix form of the Hamiltonian in the reduced basis is given in the Appendix~\ref{appenA}.
%

%
%

Therefore, we restrict all the computations  in the reduced  Hilbert space, spanned by  the above basis states.
The above Hamiltonian is easy to handle and can be  exactly  diagonalized, which yields following spectrum of eigenvalues given by   $E_1=- \omega , E_2=E_3= \omega,E_4 = \omega -\sqrt{2 g^2+\kappa^2}, E_5 =\omega +\sqrt{2 g^2+\kappa^2}, E_6= \omega  + \lambda_{+}, E_7= \omega  - \lambda _+, E_8= \omega  + \lambda _-, E_9= \omega  - \lambda _-$, where $\lambda_{\pm}$'s are given by
\begin{eqnarray}
\lambda _{\pm}&=& \frac{1}{\sqrt{2}}\sqrt{6 g^2+3 k^2\pm\sqrt{4 g^4+4 \left(5+4 \sqrt{2}\right) g^2 k^2+k^4}}. \nonumber\\
\end{eqnarray}

We  start with the following initial state characterizing the initial atom-cavity system.
\begin{eqnarray}
|\psi_{int}\rangle=(\cos{\alpha|\uparrow\uparrow 00\rangle+\sin{\alpha}|\downarrow \downarrow 00\rangle}),
\label{psi_initial}
\end{eqnarray}
Now due to the action of the time-evolution operator generated by $\mathcal{H}'$,  the initially entangled  state $|\psi_{int}\rangle$ evolves to the  final state $|\psi_f\rangle$,  which can be expressed  using the reduced basis mentioned above, in the following way $|\psi_{f}\rangle =\sum_i c_i|v_i\rangle$.
In order to  obtain the reduced state $\rho$, characterizing the density matrix of the two-atoms,  one needs to trace out the degrees of freedom associated with two cavities from  the time-evolved state, $|\psi_f\rangle$. This  would lead to the following expression of the reduced density matrix, $\rho$:

\begin{figure}[h!]
\includegraphics[scale=.32]{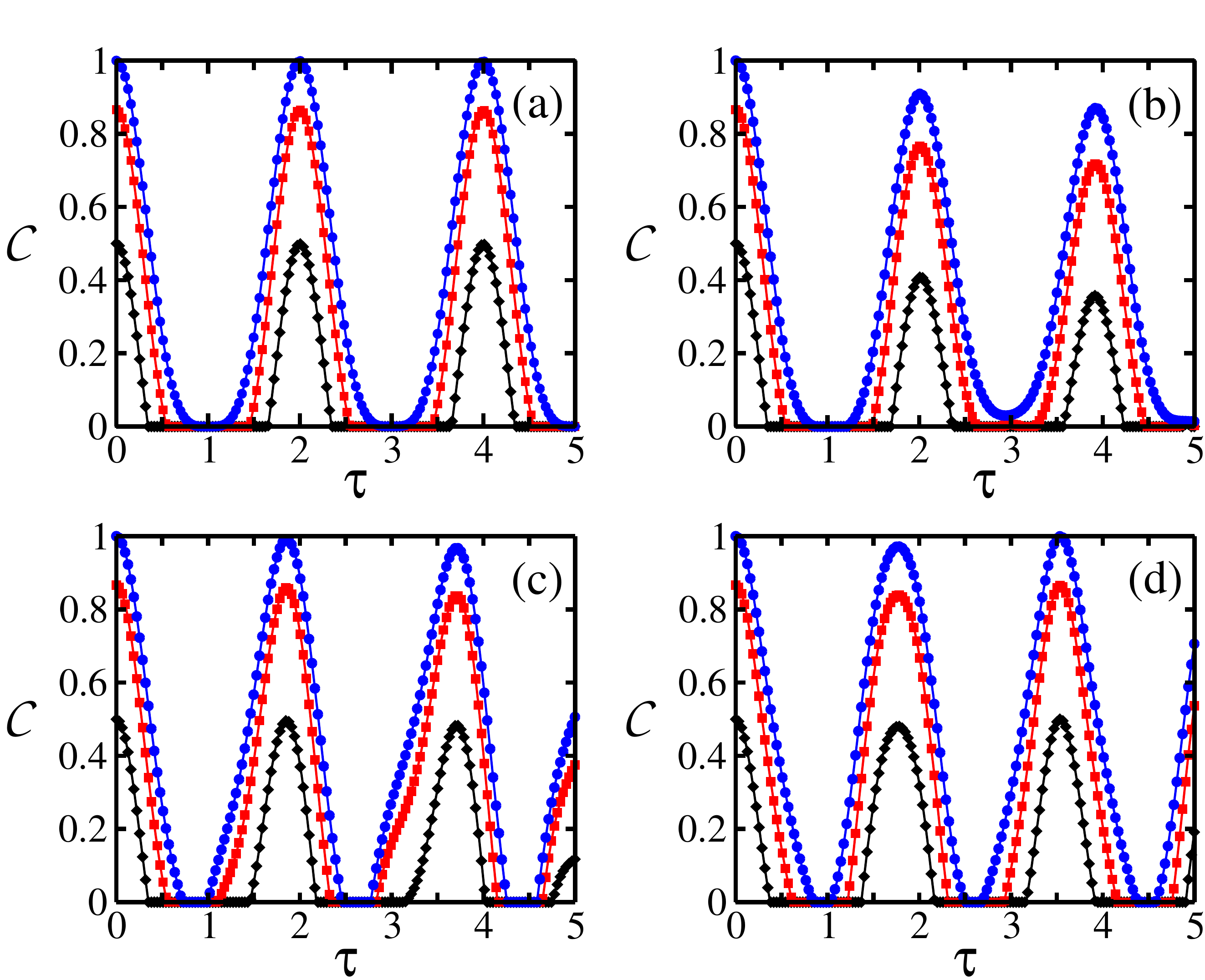}
\caption{(Color online.) Entanglement dynamics of the two-atom reduced density matrix,  $\rho$, obtained by tracing out the cavity part in the time evolved state $|\psi_f\rangle$. Each of the colors in the plots corresponds to a particular choice of the initial state in Eq.~(\ref{psi_initial}), for $\alpha=\frac{\pi}{12}$ (black-diamond), $\alpha=\frac{\pi}{6}$ (red-square), {and} $\alpha=\frac{\pi}{4}$ (blue-circle). Here we consider, $\omega = \nu = 1, g = 0.1$, (a) $\kappa = 0.001$ and (b) $\kappa = 0.01$, in the $\kappa << g$ limit, which exhibits qualitatively similar behavior of entanglement as obtained in the case of double JC model with isolated cavities. In the $k \approx g$ limit, we have (c) $\kappa = 0.1$, and (d) $\kappa = 0.2$,  where the cavity-cavity and atom-cavity  interactions are comparable leading to shortening of the time period over which ESD occurs before revival. }
\label{sudden_death_cavity_cavity}
\end{figure}

\begin{equation}
\rho =\left(
\begin{array}{cccc}
r_{11} & 0 & 0 &r_{14} \\
0 & r_{22} &0 & 0 \\
0 &0 & r_{33} & 0 \\
r_{41} & 0 & 0 & r_{44} \\
\end{array}
\right),
\label{rho}
\end{equation}
 where $c_i$'s and $r_{ij}$'s are non-trivial    functions of the  system parameters $\omega,~\nu,~g,~\kappa$, which  are explicitly derived in  the Appendix~\ref{appenB} and \ref{appenC}, respectively.

Considering   the above form of the two-site reduced density matrix $\rho$, obtained from the final time-evolved state, $|\psi_f\rangle$,  we are now equipped with the resources  required to study the effects of the cavity-cavity interaction term  on the dynamics of the entanglement between the two atoms. We  consider three different regions depending on the strength of  $\kappa$, and present a comparative study  of the behavior of  $\mathcal{C}$ in those regions.

a) \emph{$\kappa << g < \omega$} : In this limit, the strength of the cavity-cavity interaction term $\kappa$, is chosen to be much smaller than the atom-cavity interaction coefficient $g$. In Fig.~\ref{sudden_death_cavity_cavity}(a) and \ref{sudden_death_cavity_cavity}(b) we plot the variation of the entanglement as quantified by  $\mathcal{C}$ with the scaled time, $\tau = \frac{2gt}{\pi}$, for the following specification of the parameters given by  $\omega = \nu = 1, g = 0.1, \kappa = 0.001$ (Fig.~\ref{sudden_death_cavity_cavity}(a)) and $\kappa=0.01$ (Fig.~\ref{sudden_death_cavity_cavity}(b)). Now as the strength of the perturbation is very small, we  expect, the qualitative behavior of bipartite entanglement would mimic the pattern obtained in case of $\kappa=0$.  From both  the Figs.~\ref{sudden_death_cavity_cavity}(a) and \ref{sudden_death_cavity_cavity}(b), one can clearly see that in this limit, cavity-cavity interaction has no significant impact on the sudden death of entanglement and no improvement on the loss of entanglement can be observed.

b)  \emph{$\kappa \approx g < \omega$} : In this limit, the dynamics of $\mathcal{C}$ gets complicated as the strength of both the parameters,  $\kappa$ and $g$ are comparable. We plot the  dynamics of the  entanglement as quantified by $\mathcal{C}$,  in Figs.~\ref{sudden_death_cavity_cavity}(c) and \ref{sudden_death_cavity_cavity}(d) for the following specification of system parameters $\omega = \nu = 1, g = 0.1, \kappa = 0.1$ (for Fig.~\ref{sudden_death_cavity_cavity}(c)) and $\kappa = 0.2$ for Fig.~\ref{sudden_death_cavity_cavity}(d)).
Interestingly, in this case due to the effect of cavity-cavity interaction term, the time duration  over which ESD occurs, gets shortened for all value of $\alpha$, i.e.  irrespective of the amount of  entanglement in the initial state. This  essentially signifies that  the  loss of entanglement due to the perturbation induced by the cavities is  restricted upon strengthening the cavity-cavity interaction.
    \begin{figure}[h!]
\includegraphics[scale=.32]{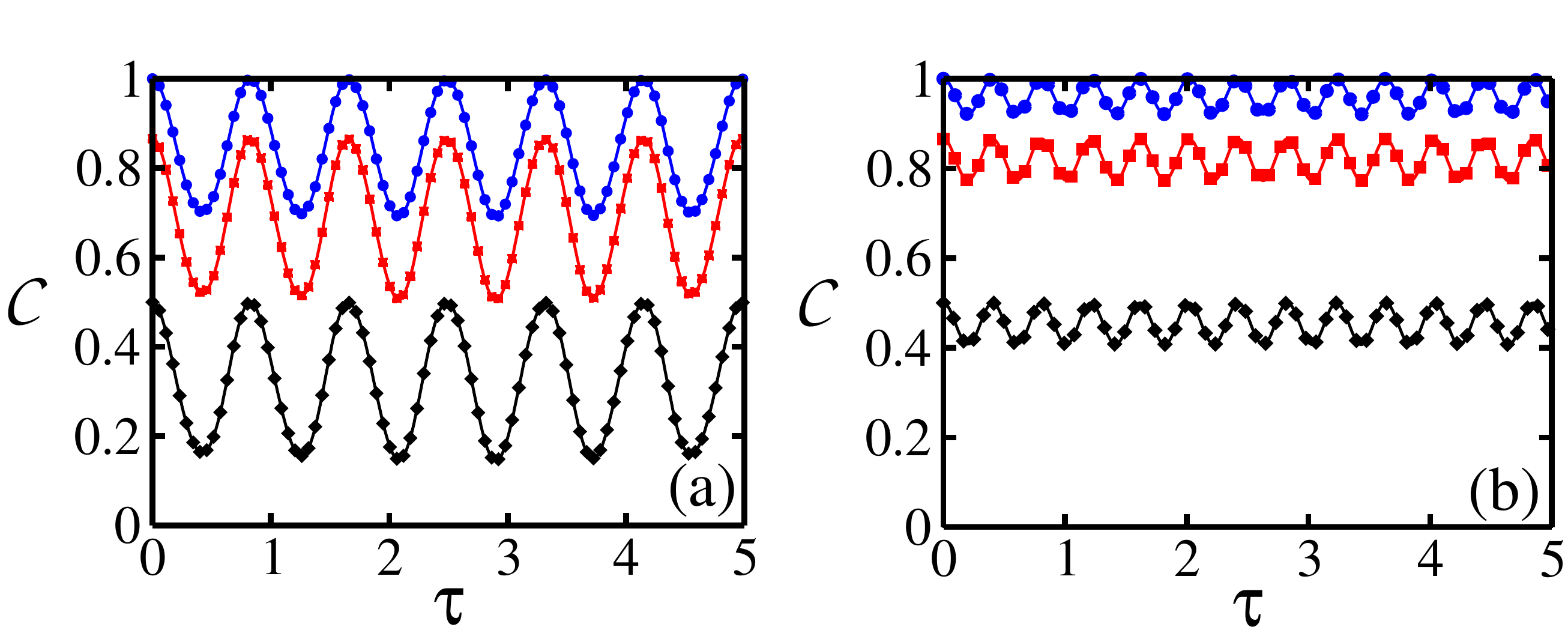}
\caption{(Color online.) Entanglement dynamics of the two-atom reduced density matrix,  $\rho$, obtained by tracing out the cavity part in the time evolved state $|\psi_f\rangle$. Each of the colors in the plots corresponds to a particular choice of the initial state in Eq.~(\ref{psi_initial}), for $\alpha=\frac{\pi}{12}$ (black-diamond), $\alpha=\frac{\pi}{6}$ (red-square), {and} $\alpha=\frac{\pi}{4}$ (blue-circle). Here we chose $\omega = \nu = 1, g = 0.1$, (a) $\kappa=0.5$ and (b) $\kappa=1.0$, in the $\kappa>>g$ limit, at which complete prevention of ESD can be achieved. Larger the value of $\kappa$, higher the  value of $\mathcal{C}$ around which system oscillates.}
\label{sudden_death_high_k}
\end{figure}

c)  \emph{$\kappa >> g$} : In this limit,  the cavity-cavity interaction dominates over the atom-cavity interaction, resulting in a complete recovery against the ESD phenomenon. Therefore,  no loss of  entanglement occurs  at any stage of the time dynamics. The variation  of the bipartite entanglement $\mathcal{C}$ with the scaled time, $\tau = \frac{2gt}{\pi}$, at large $\kappa$ limit is depicted in Fig.~\ref{sudden_death_high_k}.  Here we chose, $\omega = \nu = 1, g = 0.1, \kappa=0.5$ (for Fig.~\ref{sudden_death_high_k}(a) ) and $\kappa=1.0$  (for Fig.~\ref{sudden_death_high_k}(b)). The above findings has a remarkable significance as the entanglement between the central atoms can be protected against decoherence induced by the cavities by non-intrusively introducing photon-exchange through the cavities. This allows the coherence in the atomic system to be protected through the cavity interaction, thus completely avoiding any form of ESD.

\begin{figure}[h]
\includegraphics[scale=.315]{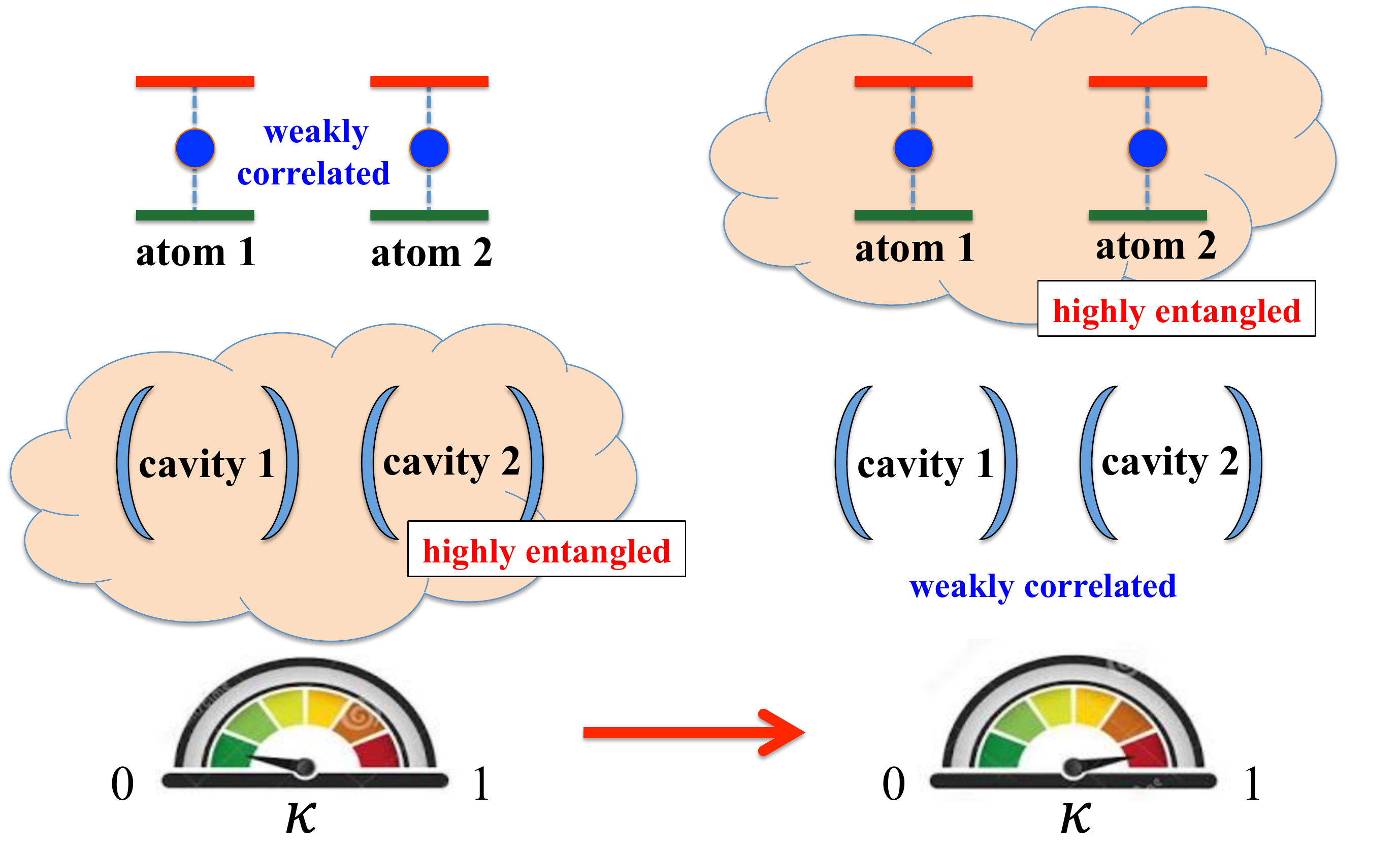}
\caption{(Color online.) Schematic diagram showing the transfer of entanglement from the atom-atom subsystem to the cavity-cavity bipartition, during the time-evolution of the double JC model. Specifically, at the limit of low $\kappa$,  the cavities remain entangled with each other, while the atomic subsystem decoheres during ESD. In the moderately large values of $\kappa$, the atoms are highly entangled with each other at all times of the evolution and the cavities almost remain weakly entangled.}
\label{schematic2}
\end{figure}

Therefore, it is clear from the above analysis that the loss of entanglement due to the cavity induced decoherence, in the double JC model can be overcome if one allows an exchange of energy through an interaction between the cavities. 
 We present a schematic diagram in  Fig.~\ref{schematic2}, which summarizes the above results in the following way: For $\kappa \rightarrow 0$, during the time-evolution, the atomic system undergoes ESD and the system shares non-zero entanglement across the cavity-cavity bipartition. Whereas,  by tuning the cavity-cavity interaction strength $\kappa$ to a moderately large value, one can protect the entanglement across the atom-atom bipartition, leaving the cavities weakly entangled to each other. 
A plausible interpretation of the behavior of entanglement in the pair of atom-cavity system is the back and forth quantum transfer of entanglement between the atom-atom and the cavity-cavity subsystems. For low values of $\kappa$, the entanglement between the two atoms is completely transferred to the pair of cavities, during the evolution, via the pair of atom-cavity quantum channels, leading to sudden death of entanglement between the atoms. The opposite process, which is the quantum transfer of entanglement from the cavities to the two atoms, leads to revival of entanglement. In contrast, for high $\kappa$, presence of considerable quantum correlations between the two-cavities inhibit such a complete transfer of entanglement from the atoms to the cavities, thus precluding the possibility of ESD across the two atoms.

%
 
\section{Effect of spin interaction}
\label{Effect of nearest neighbor spin exchange}

In order to study the entanglement dynamics in presence of the Ising interaction term, we first modify the Hamiltonian expressed in Eq.~(\ref{JC_modified}), such that
\begin{eqnarray}
\mathcal{H}'' &=& \omega \sum_{i=1}^2 \sigma^z_i+g\sum_{i=1}^2 (a^+_i\sigma^-_i+a_i\sigma^+_i) +\nu\sum_{i=1}^2 a_i^+ a_i \nonumber\\
&+& \kappa \sum_{\substack{i,j=1\\i\neq j}}^2 a^+_i a_j+J~ \sigma^z_1\otimes\sigma^z_2,
\label{JC_modified2}
\end{eqnarray}
where $J$ is the Ising coupling strength between the two atoms in the cavities. The other terms are the same as in Eq.~(\ref{JC_modified}). We constraint the cavity-cavity interaction to the weak regime, i.e. $\kappa < g$.  As in the previous case, we present all the calculations in the reduced subspace of the above Hamiltonian, spanned by the energy-conserved basis states mentioned in the previous section. 
Therefore, the final form of the time-evolved state $|\psi_{f}\rangle$, obtained from the initially entangled state $|\psi_{in}\rangle$ as expressed in Eq.~(\ref{psi_initial}), remains the same as given in the previous section, albeit with a different set of coefficients. However, the exact analytical form  of the coefficients $c_i$'s, which are now functions of the parameters $\omega, g, \kappa$, and $J$, are not analytically simple but nonetheless can be
computed numerically without much effort. The two-atom reduced density matrix, $\rho$, and its entanglement, $\mathcal{C}$, can then be estimated from $|\psi_{f}\rangle$.
\begin{figure}[t]
\includegraphics[scale=.32]{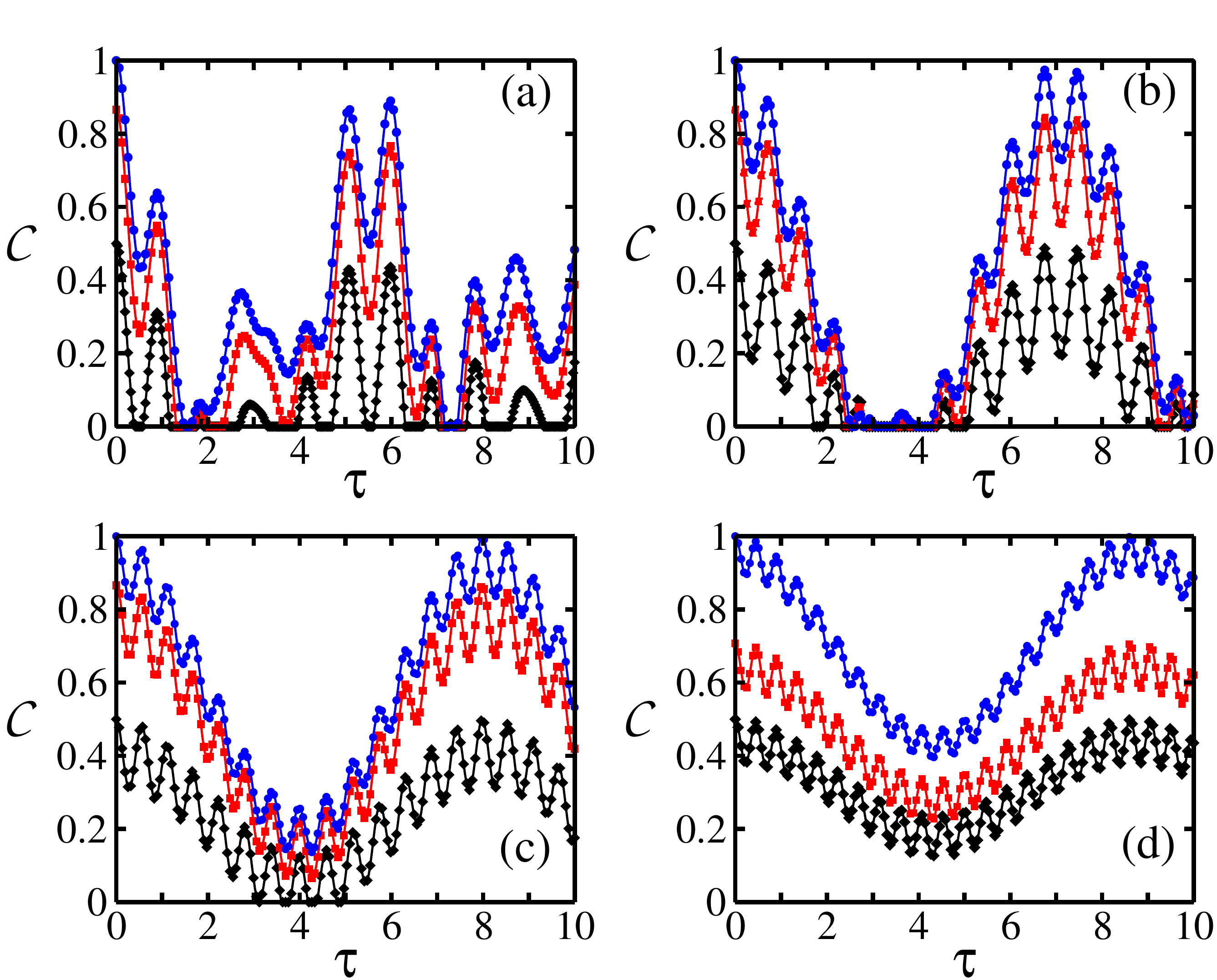}
\caption{(Color online.) Entanglement dynamics of the two-atom reduced density matrix $\rho$ governed by the  Hamiltonian $\mathcal{H}''$, given by Eq.~(\ref{JC_modified2}), for initial states with $\alpha=\frac{\pi}{12}$ (black-diamond), $\alpha=\frac{\pi}{6}$ (red-square), {and} $\alpha=\frac{\pi}{4}$ (blue-circle). Here, we consider $\omega = \nu = 1,~ g = 0.5, ~\kappa = 0.1$, with (a) $J=0.5$, (b) $J=1.0$, (c) $J=1.5$, and $J=2.0$. 
}
\label{sudden_death_ising}
\end{figure}

Figure \ref{sudden_death_ising}, shows the variation of entanglement, as quantified by $\mathcal{C}$, with scaled time, $\tau = \frac{2gt}{\pi}$, for a similar set of initial sets. From the figure, it is evident that unlike the case of cavity-cavity interactions, the presence of atomic interactions is able to effectively curb ESD, induced by cavity perturbations, only at high values of spin-spin coupling, i.e. in the $J >> g >> \kappa$ limit (see Fig.~\ref{sudden_death_ising}(d))). Even at moderately large value of the Ising coupling parameter, effects of  decoherence dominates the atomic interactions and the atomic system fails to prevent the loss of bipartite entanglement (see Figs.~\ref{sudden_death_ising}(a) - \ref{sudden_death_ising}(c)).

We next consider the case where  the two atoms interact via the spin-exchange interactions governed by the  well-known anisotropic $XY$ Hamiltonian. The double JC Hamiltonian can be modified to the following form:
\begin{eqnarray}
\mathcal{H}''&=&\omega \sum_{i=1}^2 \sigma^i_z+g\sum_{i=1}^2 (a^+_i\sigma^-_i+a_i\sigma^+_i) +\nu\sum_{i=1}^2 a_i^+ a_i \nonumber\\
&+& \kappa \sum_{\substack{i,j=1\\i\neq j}}^2 a^+_i a_j + \big(J_x~ \sigma^1_x\otimes\sigma^2_x
+ J_y~ \sigma^1_y\otimes\sigma^2_y\big),
\label{JC_modified3}
\end{eqnarray}
where $J_x$ and $J_y$ are the coupling constants along the $x$ and $y$-directions, respectively, and the anisotropy arises due to $J_x \neq J_y$.
Once more, we restrict the cavity-cavity interaction, $\kappa$ to weak values. Numerical studies of the reduced atomic density matrix, $\rho$, show that inclusion of the $XY$ interaction term in $\mathcal{H}''$ has no systematic effects on the entanglement dynamics. We encounter initial states, for which ESD can not be overcome for any value of the parameters $J_x$ and $J_y$ (For instance, the initial state with $\alpha=\frac{\pi}{6}$ and $\mathcal{H}''$ with $\omega=\nu=1,~g=0.5,~\kappa=0.001,~J_x=1.95,~J_y=0.05$).
Hence, starting from an unknown initially entangled state, it is in general difficult to find a choice of parameters for which the state prevents sudden death of entanglement.
%

%
%
%
%

\section{Discussion}
\label{conclusion}

In this work, we considered  a double JC Hamiltonian, consisting of two spatially separated atoms, each inside an, initially non-interacting, lossless cavity. We investigated the entanglement dynamics associated with the system with the specific aim  at countering the loss of entanglement and sudden death in the time-evolved two-atom state. Motivated by contemporary studies on strongly correlated dynamics in arrays of optical cavities \cite{qptlight,oztop}, we consider a photon-exchange interaction between the two cavities in the double JC model. This allows us to introduce a non-intrusive, cavity-mediated interaction between the spatially separated atoms. We observe that in the limit of low cavity-cavity coupling, $\kappa < g$, no significant improvement on the decoherence of entanglement between the atoms is noticed during the course of evolution, i.e., the system still undergoes ESD. However, as the coupling strength is moderately increased ($\kappa \gtrsim 0.5 > g$), a more robust response of bipartite entanglement can be observed, which results in complete preclusion of the ESD phenomenon during the atom-cavity evolution.
We also verify whether the effect of a more direct interaction between the atoms can mitigate the decoherence effects induced by the cavity. By setting cavity interactions in the weak regime and including spin-exchange interaction, pertaining to the Ising and $XY$ class of Hamiltonians, we observe that the  effects are not very potent, nor conclusive, as compared to the non-intrusive cavity hopping mechanism.
 This shows that the cavity-mediated interaction is a more robust approach to countering the decoherence of entanglement in the archetypal double JC model. With the existence of experimental apparatus to implement photon-exchange between cavities, such as in photonic band gap structures in NV centers of diamonds \cite{nv} and microwave strip line resonators \cite{strip}, along with developments in Rydberg atoms in superconducting cavities \cite{ry}, the results presented could provide an important direction in overcoming loss of coherence in quantum protocols.


%

\begin{widetext}
\appendix
\section{ Matrix form of $\mathcal{H}'$ in the energy conserving basis $\{|v_i\rangle\}$ }
\label{appenA}
Here we present the matrix form of the generalized JC Hamiltonian expressed in Eq.~(\ref{JC_modified}), using the energy conserving basis of the total atom-cavity system
$\{|\hspace{-0.1cm}\uparrow\uparrow\hspace{-0.1cm}00\rangle,  |
\hspace{-0.1cm}\uparrow\downarrow\hspace{-0.1cm}10\rangle,|
\hspace{-0.1cm}\uparrow\downarrow\hspace{-0.1cm}01\rangle,|
\hspace{-0.1cm}\downarrow\uparrow\hspace{-0.1cm}10\rangle,|
\hspace{-0.1cm}\downarrow\uparrow\hspace{-0.1cm}01\rangle,|
\hspace{-0.1cm}\downarrow\downarrow\hspace{-0.1cm}20\rangle,|
\hspace{-0.1cm}\downarrow\downarrow\hspace{-0.1cm}11\rangle,|
\hspace{-0.1cm}\downarrow\downarrow\hspace{-0.1cm}02\rangle,|
\hspace{-0.1cm}\downarrow\downarrow\hspace{-0.1cm}00\rangle$\}, in the following way
\begin{equation}
\label{matrix_form}
\mathcal{H}' =\left(
\begin{array}{ccccccccc}
-\omega  & 0 & 0 & 0 & 0 & 0 & 0 & 0 & 0 \\
0 & \omega  & k & 0 & \sqrt{2} g & 0 & 0 & 0 & 0 \\
0 & k & \omega  & k & 0 & g & g & 0 & 0 \\
0 & 0 & k & \omega  & 0 & 0 & 0 & \sqrt{2} g & 0 \\
0 & \sqrt{2} g & 0 & 0 & \omega  & k & 0 & 0 & 0 \\
0 & 0 & g & 0 & k & \omega  & 0 & 0 & g \\
0 & 0 & g & 0 & 0 & 0 & \omega  & k & g \\
0 & 0 & 0 & \sqrt{2} g & 0 & 0 & k & \omega  & 0 \\
0 & 0 & 0 & 0 & 0 & g & g & 0 & \omega  \\
\end{array}
\right).
\\ \\ \\
\end{equation}
\section{  Mathematical expression of the coefficients $c_i$'s in time evolved  state $|\psi_f\rangle$}
\label{appenB}

Here we present the mathematical expressions of the coefficients  $c_i$'s, present in the expression of $|\psi_f\rangle$.
\begin{eqnarray}
c_1 &=& e^{i t \omega} \sin{\alpha},\nonumber\\ \nonumber\\
c_2 &=&c_4 = i \left(1-\sqrt{2}\right) e^{-i t \omega} g^2 k \cos{\alpha} \Big(N_2 \sin({\lambda_{-}} t)\nonumber\\&+&M_2 \sin({\lambda_{+}} t
\Big)\nonumber, \\ \nonumber\\
c_3 &=& \Big(\left(e^{- i E_9 t}+e^{-i E_8 t}\right) \frac{1}{N_3}-\left(e^{- i E_6 t} +e^{- i E_7 t}
\right)\frac{1}{M_3}\nonumber\\&+&\frac{e^{-i t \omega} \left(\sqrt{2} g^2k^2- 2 g^4\right)}{(2 \left(1-\sqrt{2}\right)g^2 k^2  + 4g^4 + k^4)}\Big)\cos{\alpha} \nonumber,\\ \nonumber\\
c_5 &=& c_8 = \Bigg(\left((e^{- i E_8t} +e^{- i E_9t} )\frac{1}{N_4}-(e^{-i E_6t}+e^{- i E_7t} )\frac{1}{M_4}\right)\nonumber\\&+&\frac{e^{-it \omega} g k (\sqrt{2} g^2-k^2)}{ (2 (1-\sqrt{2})g^2 k^2 + 4g^4 + k^4)}\Bigg)\cos\alpha, \nonumber \\ \nonumber\\
c_6 &=& c_7 =\left(\left( e^{- i t E_9} - e^{- i t E_8}\right)\frac{\lambda_+}{\gamma _+}+\left( e^{- i t E_6}-e^{- i t E_7} \right)\frac{\lambda_-}{\gamma_-}\right) \nonumber\\&&\sqrt{2}g^3 k^2\cos\alpha\nonumber,\\
c_9&=&\Bigg((-( e^{- i E_8 t}+e^{- i E_9 t})\frac{1}{\gamma _+}+( e^{- i E_6t} + e^
{- i E_7t})\frac{1}{\gamma_-})N_5\nonumber
+\frac{e^{ it\omega}\left(-\sqrt{2} g^2+k^2\right)^2}{\left( \left(2-2^{\frac{3}{2}}\right)g^2 k^2 + 4g^4 + k^4\right)}\Bigg)\cos\alpha,\nonumber\\ \nonumber\\
\text{with}\nonumber\\
N_2&=& \frac{\left(4 g^4 +k^4+4 g^2k^2 \left(5+4 \sqrt{2}\right) +3\delta \left(2g^2+ k^2 \right) \right)\lambda _-}{2 \left( \left(-3\sqrt{2}+4\right)
\left(16g^8+k^8\right)+ 2g^2k^2\left( \left(-5\sqrt{2}+2\right) \left(4g^4+k^4\right)-8 \left(1-\sqrt{2}\right) g^2 k^2\right)\right)}, \nonumber\\ \nonumber\\
N_3&=& \left(\sqrt{2}\gamma _+\right)/\left(g^2 k^2 \left(-2 \sqrt{2} g^2+\left(-2+\sqrt{2}\right) \left(k^2+\delta \right)\right)\right)\nonumber,\\ \nonumber\\
N_4 &=&\left(\sqrt{2}\gamma _+\right)/\left(g^3 k \left(\left(-1+\sqrt{2}\right) \left(2g^2-\delta \right)-\left(1+\sqrt{2}\right) k^2\right)\right),~~ N_5 = 2^{\frac{3}{2}}(g^2 k)^2, \nonumber\\ \nonumber\\
 M_2&=& \frac{\left(4 g^4 +k^4+4 g^2k^2 \left(5+4 \sqrt{2}\right) -3\delta \left(2g^2+ k^2 \right) \right)\lambda _+}{2 \left( \left(-3\sqrt{2}+4\right)
\left(16g^8+k^8\right)+ 2g^2k^2\left( \left(-5\sqrt{2}+2\right) \left(4g^4+k^4\right)-8 \left(1-\sqrt{2}\right) g^2 k^2\right)\right)}\nonumber, \\ \nonumber\\
M_3&=& \left(\sqrt{2}\gamma _-\right)/\left(g^2 k^2 \left(-2 \sqrt{2} g^2-\left(-2+\sqrt{2}\right) \left(-k^2+\delta \right)\right)\right)\nonumber,\\ \nonumber\\
M_4&=& \left(\sqrt{2}\gamma _-\right)/\left(g^3 k \left(\left(-1+\sqrt{2}\right) \left(2g^2+\delta \right)-\left(1+\sqrt{2}\right) k^2\right)\right),~~~~\text{and}\nonumber\\
\gamma _{\pm}&=&  \pm \left(4-3 \sqrt{2}\right)\left(4 g^6+ k^4 \left(k^2\pm\delta  \right)\right)+g^2 k^2 \left(\left(\pm(-12+\sqrt{2})\right) k^2+\left(-4+\sqrt{2}\right)\nonumber
\delta  \right)\\
&+& g^4 \left(\mp8 \sqrt{2} k^2+2 \left(-4+3 \sqrt{2}\right)\delta  \right), \nonumber\\ \nonumber\\
\delta &=& \sqrt{4 g^4+4 \left(5+4 \sqrt{2}\right) g^2 k^2+k^4}.
\end{eqnarray}

 \section{Mathematical expression of the elements of the reduced density matrix $\rho$ }
 \label{appenC}
 To  obtain the reduced state $\rho$, characterizing the density matrix of the two-atoms,  one needs to trace out the degrees of freedom associated with two cavities from  the time-evolved state, $|\psi_f\rangle$. This  would lead to the following expression of the reduced density matrix, $\rho$:
\begin{equation}
\rho =\left(
\begin{array}{cccc}
r_{11} & 0 & 0 &r_{14} \\
0 & r_{22} &0 & 0 \\
0 &0 & r_{33} & 0 \\
r_{41} & 0 & 0 & r_{44} \\
\end{array}
\right),
\label{rho}
\vspace{.5cm}
\end{equation}

where $r_{ij}$'s are given by

\begin{eqnarray}
r_{11}&=& (\sin\alpha)^2 + 8 \left(3-2 \sqrt{2}\right) \cos\alpha^2 \left(\frac{1}{A_1}\left(k^2-\eta_+ \right)
\sin\left( \lambda _-t\right) \lambda _-+\frac{1}{A_2} \left(k^2-\eta_- \right) \sin\left(\lambda _+t\right) \lambda _+\right)^2+\cos\alpha^2
\nonumber\\&\times&\left(\frac{2 g^4-\sqrt{2} g^2 k^2}{4 g^4+A_52  g^2 k^2+k^4}+\frac{2 \sqrt{2} \left(- 2 g^2+A_5 \left(k^2-\delta
\right)\right) \cos( \lambda _-t)}{A_3}+\frac{2\sqrt{2} \left(- 2 g^2+A_5 \left(k^2+\delta \right)\right)
cos\left( \lambda _+t\right)}{A_4}\right)^2,\nonumber\\
r_{14}& =& r_{41} = \left(\frac{ \left(-\sqrt{2} g^2+k^2\right)^2}{\left(4 g^4+2 A_5 g^2 k^2+k^4\right)}+\frac{2}{A_3}
\cos (\lambda _-t)+\frac{2}{A_4} \cos( \lambda _+t)\right)\cos\alpha~\sin\alpha ~e^{i 2 t\omega}\nonumber,\\
r_{22} &=& g^2 \cos\alpha ^2\left(64k^2\left(\frac{1}{A_1} \sin( \lambda _-t) \lambda _-+\frac{1}{A_2}
\sin( \lambda _+t) \lambda _+\right)^2+ \left(2 A_6 k \left(\frac{1}{A_3} \cos( \lambda
_-t)+\frac{1}{A_4} \cos( \lambda _+t)\right)\nonumber +A_7\right)^2\right)\nonumber,\nonumber\\
r_{33} &=& r_{22},~~ r_{44} = \cos\alpha^2 \left(1-\frac{2 g^2 \left(g^2+k^2\right)}{4 g^4+2 A_5 g^2 k^2+k^4}+\frac{2\cos(
\lambda _-t)}{A_3}+\frac{2 \cos(\lambda _+t)}{A_4}\right)^2,~\text{with}\nonumber\\
A_1& =& \frac{1}{2}g^2 k \left(16+\frac{2 \left(-A_5\eta_+-A_6 k^2\right)^2}{g^2 k^2}+\frac{\left(2
\sqrt{2} g^2+\sqrt{2}A_5 \left(-k^2+\delta \right)\right)^2}{g^4}+\frac{8 \lambda_-^2}{g^2}+\frac{\left(3-2 \sqrt{2}\right) \left(\eta_+-k^2 \right)^2 \lambda_-^2}{g^4 k^2}\right), \nonumber\\
A_2& =& \frac{1}{2} g^2 k \left(16+\frac{2 \left(A_5 \eta_-+A_6 k^2\right)^2}{g^2
k^2}+\frac{\left(-2 \sqrt{2} g^2+\sqrt{2}A_5 \left(k^2+\delta \right)\right)^2}{g^4}+\frac{8 \lambda _+^2}{g^2}+\frac{\left(3-2 \sqrt{2}\right)
\left(k^2-\eta_- \right)^2 \lambda_+^2}{g^4 k^2}\right),\nonumber\\
A_3& =&\frac{A_1}{2k},
A_4  =\frac{A_2}{2k},
A_5=-\sqrt{2}+1,~ A_6=1+\sqrt{2},\nonumber \\A_7&=& \frac{2}{k}  A_5\left(\frac{1}{A_4} \eta_- \cos(\lambda _+t)+\frac{1}{A_3}\eta_+ \cos( \lambda _-t)\right)+\frac{\sqrt{2} g^2 k-k^3}{4 g^4-2 A_5 g^2 k^2+k^4},
\eta_{\pm}=2g^2\pm\delta .
\end{eqnarray}

\end{widetext}

 \end{document}